\documentclass[prl,aps,twocolumn, floatfix, superscriptaddress,showpacs]{revtex4}
\usepackage{amsmath,bm,graphicx}
\usepackage{amssymb}
\usepackage{amsfonts}
\newcommand{\w}{\omega}
\newcommand{\nn}{N}

\newcommand{\dd}[2]{\frac{d #1}{d #2}}

\begin{document}

\title{Aggregation--Fragmentation Processes and Wave Kinetics}

\author{Colm Connaughton}
\email{connaughtonc@gmail.com}
\affiliation {Mathematics Institute and Centre for Complexity Science, University of Warwick, Gibbet Hill
Road, Coventry CV4 7AL, UK}
\author{P. L. Krapivsky}
\affiliation {Department of Physics, Boston University, Boston, Massachusetts 02215, USA}
\affiliation{LPT -- IRSAMC, CNRS, Universit\'e de Toulouse, 31062 Toulouse, France}
\date{\today}

\begin{abstract}
There is a formal correspondence between the isotropic 3-wave kinetic 
equation and the rate equations for a non-linear fragmentation--aggregation
process. We exploit this correspondence
to study analytically the time evolution of the wave frequency power spectrum.
Specifically, we analyzed a 3-wave turbulence in which the wave interaction 
kernel is a constant. We consider both forced and decaying
turbulence. In the forced case, the scaling function diverges as 
$x^{-3/2}$ as expected from Kolmogorov--Zakharov theory. In the 
decaying case, the scaling function exhibits non-trivial, and hitherto 
unexpected,  divergence with both algebraic and logarithmic spectral exponents
which we calculate. This divergence leads to non-trivial decay laws for the
total wave action and the number of primary waves. All theoretical predictions
are verified with high quality numerical simulations of the 3-wave kinetic 
equation.
\end{abstract}

\pacs{47.35.-i, 82.20.-w,  94.05.Lk}
\maketitle

Wave turbulence is a theory of the statistical evolution of ensembles of 
weakly nonlinear dispersive waves. It has been applied to capillary waves on
fluid interfaces, gravity waves on the ocean, acoustic turbulence and various
special limits of plasma and geophysical turbulence. (For a review of the theory
see \cite{ZLF92}; for a summary of applications see \cite{CNN03}.) The 
key feature is the fact that weak nonlinearity permits the consistent 
derivation \cite{NNB01} of a wave kinetic equation describing 
the time evolution of the frequency power spectrum, $N_\w(t)$.  When sources
and sinks of energy, widely separated in frequency, are added to the wave
kinetic equation, it can be shown to have exact stationary solutions
corresponding to a cascade of energy through frequency space from the source
to the sink. The cascade solution is known as the Kolmogorov-Zakharov (K-Z)
spectrum; it describes an intrinsically non-equilibrium state of the wave
field. Everything is known about the stationary K-Z spectra, their scaling 
exponents, locality  and stability. 
By contrast, very little is known about the time-dependent solutions of
the wave kinetic equation. A basic scaling theory of the development of the
stationary state in the case of forced wave turbulence was provided in
\cite{FS1991} although numerical investigations \cite{GNNP2000,CNP03} have 
suggested that there are unexplained dynamical scaling anomalies in many cases.
Almost nothing is known about time-dependent solutions in the case
of decaying turbulence where an initial specrum is allowed to decay
in the absence of external forcing. In this Letter we take the first steps
to remedy this.

The subject of aggregation--fragmentation kinetics, having its origins in
theoretical chemistry has, at first sight, rather little to do with waves
or turbulence. This field concerns itself with the statistical mechanics of
ensembles of particles which aggregate or fragment upon contact. The principal
quantity of interest is the  particle size distribution, $n_i(t)$, denoting
the density of clusters of mass $i$ at time $t$. It satisfies a kinetic
equation, which, in the case of pure aggregation, is the well-known 
Smoluchowski coagulation equation \cite{SMO1917}. For a review of pure
aggregation processes see \cite{DRA72}. If clusters also break up,
additional terms may be added to the Smoluchowski equation to
take this into account. See \cite{RED1990} for a review of fragmentation.
In aggregation--fragmentation kinetics, in strong contrast with wave
kinetics, almost all theoretical effort has historically been focused on 
determining the time evolution of $n_i(t)$ from the underlying
kinetic equation. As a result,  a  comprehensive scaling theory
of the solutions of the Smoluchowski equation has been constructed (see
\cite{LEY2003} for a review). Although there is a conceptual analogy 
\cite{STEP1989} between energy transfer between scales in turbulence and
mass transfer between clusters in aggregation, it is only recently that this
analogy has been made quantitatively useful. 
Concepts and techniques from turbulence have proven useful in analysing
certain aspects of aggregation problems \cite{CRZ2004,CRZ2005,CRZ2006}. 
Furthermore, it has been shown \cite{CON2009} that, in the case of
isotropic wave turbulence with quadratic nonlinearity, the  wave kinetic 
equation can be rewritten as a set of
rate equations for a aggregation-fragmentation process with an unusual
nonlinear fragmentation mechanism. This correspondence opens the door for
the transfer of ideas and techniques from aggregation--fragmentation kinetics to the
context of wave turbulence which will hopefully start to fill in the gap
in knowledge of time-dependent solutions of the wave kinetic equations
alluded to already. Furthermore, this correspondence opens up a new 
set of problems within aggregation--fragmentation kinetics. This Letter
contains some opening explorations in this direction.

It was shown in \cite{CON2009} that resonant interactions between waves lead 
to forward transfer of energy
between frequencies which looks like an aggregation process:
$(i)\oplus (j)\to (i+j)$.
Back-scatter of energy leads to a fragmentation process
$(i)\oplus (i+j)\to (i)\oplus (i) \oplus (j)$.
This fragmentation mechanism  is unusual. It is non-linear while 
typically  \cite{RED1990} the fragmentation mechanism is linear: $(i+j)\to (i) \oplus (j)$. 
Nonlinear collision-controlled fragmentation processes have been studied in 
the past (see \cite{KB2003} and references therein). While they are somewhat 
similar to the above rule, this model asserts that only the larger 
particle breaks and it happens according to a rather special rule.
Our goal here is to apply ideas and techniques developed in studies 
of aggregation and fragmentation to wave turbulence. We want to examine  
fundamental aspects and have, therefore, limited ourselves to the simplest 
possible situation where the wave interaction kernels are constant.
The dynamical problem is already non-trivial at this level.
The resulting kinetic equation can be reduced (see \cite{CON2009} for details),
in the discrete case \footnote{Discreteness means that 
frequencies are multiples of the primary frequency. The 
wave-action distribution is then $N_\w(t)$ with $\w=1,2,\ldots$. In the scaling 
limit, there is no difference between discrete and continuous. }, to:
\begin{eqnarray}
\label{eq-3WKE}
\frac{d N_\w}{dt} &=& J\,\delta_{\w\,1} + 
\frac{1}{2}\sum_{\w_1+\w_2=\w}N_{\w_1} N_{\w_2} - N_\w\sum_{\w_1\geq 1}N_{\w_1}\\
\nonumber&-&N_\w\sum_{\w_1<\w}N_{\w_1} + N_\w\sum_{\w_1>\w}N_{\w_1} +\sum_{\w_1\geq 1}N_{\w_1} N_{\w+\w_1}
\end{eqnarray}
where $J$ is the energy injection rate and $N_\w$ is the frequency space wave action.
The total wave action is $N(t)=\sum_{\w\geq 1} N_\w(t)$. In the decay case ($J=0$), it satisfies the 
equation (found by summing  Eqs.~\eqref{eq-3WKE})
\begin{equation}
\label{eq-density}
\frac{d N}{dt} = - \frac{1}{2}\sum_{\w\geq 1}N_\w^2.
\end{equation}
The primary waves (monomers) evolve according to
\begin{equation}
\label{eq-monomers}
\frac{d N_1}{dt} = - N_1^2 + \sum_{\w\geq 1}N_\w N_{\w+1}.
\end{equation}

We assume the {\em scaling hypothesis}: there exists a typical
scale, $s(t)$, such that  $N_\w(t)$ is asymtotically of the form
\begin{equation}
\label{eq-scalingHypothesis}
\nn_\w(t) = s^{a} F(\w/s).
\end{equation}
Given this hypothesis, it follows from Eq.~(\ref{eq-3WKE}) that
\begin{equation}
\label{eq-sEvolution}
\dd{s}{t} = s^{a+2}
\end{equation}
while $F(x)$ must satisfy a complicated integro-differential equation.
The scale $s(t)$ is defined as a ratio of moments:
\begin{equation}
\label{eq-typicalScale}
s(t) = \frac{M_2(t)}{M_1(t)} \hspace{0.5cm}M_n(t)=\int_0^\infty w^n \nn_\w(t)\,d\w
\end{equation}
Often the scaling function, $F(x)$, diverges at small $x$:
\begin{equation}
\label{eq-tau}
F(x) \sim A\, x^{-\tau}  \hspace{0.25cm}\mbox{as $x\to 0$.}
\end{equation}
The exponent $\tau$ is the wave spectrum exponent or polydispersity exponent.
The shape of the frequency power spectrum for large time is determined by 
the small $x$ behaviour of the scaling function, $F(x)$.
In aggregation problems this divergence has been often encountered and the 
$\tau$ has proven to be difficult to determine \cite{LEY2003,CS1997}; 
in some seemingly simple models the exponent $\tau$ remains unknown. An 
important lesson from this work is that one should be particularly careful  
when $\tau\geq 1$.

\begin{figure}[htbp]
\centering
\includegraphics[width=0.4\textwidth]{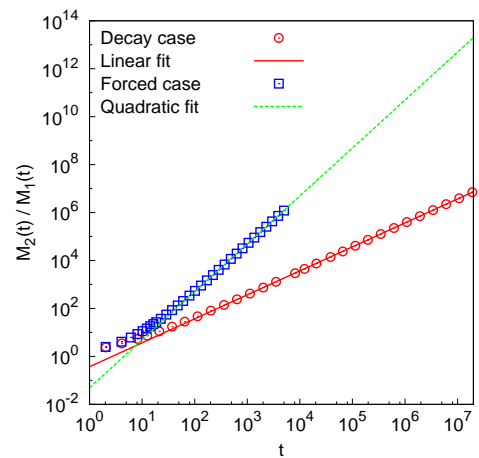}
\caption{Time evolution of typical frequency, Eq.~(\ref{eq-typicalScale}),
for forced and decaying turbulence. The dashed lines correspond to the 
theoretical predictions provided by Eq.~(\ref{eq-forcedScaling}) and 
Eq.~(\ref{eq-decayScaling}) respectively}
\label{fig-typicalFreq}
\end{figure}

This is a finite capacity system so there is no dissipative anomaly / gelation 
transition. Energy is therefore conserved for all time by the wave
interactions. For the forced case, the total energy then grows linearly
in time since we are injecting energy at a constant rate. Thus $M_1 \sim t$
(we take $J=1$). The scaling hypothesis, Eq.~(\ref{eq-scalingHypothesis}),
 then implies that $a=-\frac{3}{2}$
and subsequently solving Eq.~(\ref{eq-sEvolution}) suggests that 
$s(t)\sim r_0\,t^2$ for some constant, $r_0$.  We then expect the scaling
\begin{equation}
\label{eq-forcedScaling}
\nn_\w(t) \sim s^{-3/2}\,F(\w/s)\hspace{0.25cm}\mbox{with $s\sim r_0\,t^2$.}
\end{equation}
On the other
hand, for the decaying turbulence energy is conserved , $M_1(t)=1$. The 
scaling hypothesis then implies that $a=-2$
and solving Eq.~(\ref{eq-sEvolution}) gives $s(t)\sim s_0\,t$ for some
constant, $s_0$. We then expect the scaling
\begin{equation}
\label{eq-decayScaling}
\nn_\w(t) \sim s^{-2}\,F(\w/s)\hspace{0.25cm}\mbox{with $s\sim s_0\,t$.}
\end{equation}
These predictions for the growth rate of $s(t)$,
which are based solely on the assumption of scaling and the absence of a
dissipative anomaly, are verfied numerically in Fig.~\ref{fig-typicalFreq}.
All numerics have been 
done using the algorithm described in \cite{CON2009}.

\begin{figure}[htbp]
\includegraphics[width=0.4\textwidth]{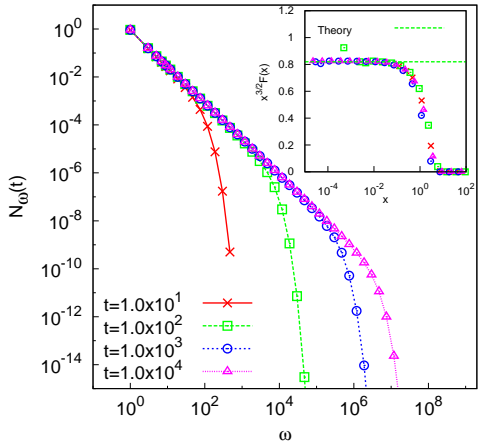}
\caption{Time evolution of wave spectrum in the forced case. The main panel
shows snapshots of $\nn_\w(t)$ at a succession of times. The inset shows the
same data collapsed according to the scaling in Eq.~(\ref{eq-forcedScaling}).
The collapsed data has been compensated by $x^{3/2}$ in accordance with 
Eq.~(\ref{eq-forcedFAsymptotics}) and shows a plateau with the theoretically 
predicted amplitude.}
\label{fig-scalingForced}
\end{figure}

For forced turbulence, we expect that
the frequency spectrum should become stationary for large times. This is an
additional piece of information which allows us to fix the spectral exponent.
Requiring that Eq.~(\ref{eq-forcedScaling}) is independent of $t$ for small
$\w$ selects $\tau=3/2$. This corresponds to
the K-Z exponent for this model \cite{CON2009}. Furthermore, the 
corresponding K-Z constant can be computed exactly for this 
model \cite{CON2009} so that we we obtain asymptotic behaviour of the scaling function
\begin{equation}
\label{eq-forcedFAsymptotics}
F(x) \sim \frac{x^{-\frac{3}{2}}}{2\sqrt{\pi-4\ln 2}}\hspace{0.25cm}\mbox{as $x\to 0$}.
\end{equation}
This prediction, and the scaling behaviour, Eq.~(\ref{eq-forcedScaling}) 
are verified explicitly from the numerical data in Fig.~\ref{fig-scalingForced}.

\begin{figure}[htbp]
\includegraphics[width=0.4\textwidth]{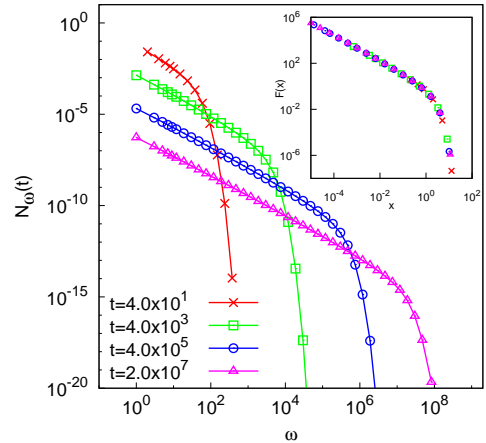}
\caption{Time evolution of wave spectrum in the decay case. The main panel
shows snapshots of $\nn_\w(t)$ at a succession of times. The inset shows same 
data collapsed according to the scaling in Eq.~(\ref{eq-decayScaling}).}
\label{fig-scalingDecay}
\end{figure}

Let us now turn to the decay case. Figure \ref{fig-scalingDecay} presents
numerical simulations of the decay of a monochromatic initial spectrum 
$\nn_\w(0)=\delta_{\w\,1}$ and verifies the scaling behaviour expected
from Eq.~\eqref{eq-decayScaling}. The essential difference from the
forced case is that we no longer have the additional constraint provided
by stationarity which allowed us to easily determine the spectral
exponent $\tau$. We must return to the original kinetic equation.

\begin{figure}[htbp]
\includegraphics[width=0.4\textwidth]{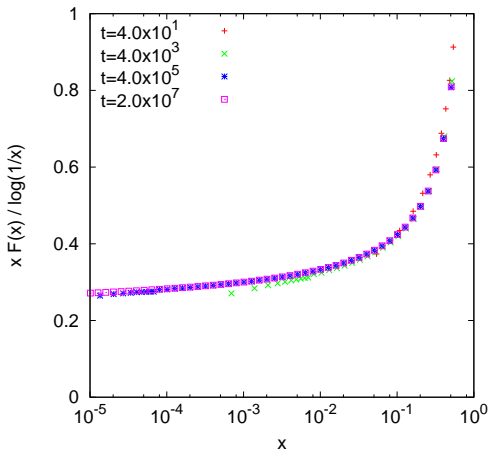}
\caption{Scaling function, $F(x)$, compensated for the theoretical small x divergence, $x^{-1}\,\ln(1/x)$, expected from Eq.~(\ref{eq-decayFAsymptotics}).}
\label{fig-compensatedScaling}
\end{figure}

Guided by our result for the forced cased, let us {\em presume} that the wave 
spectrum diverges as $x\to 0$ in accordance with Eq.~(\ref{eq-tau}). 
Substitution of Eq.~(\ref{eq-scalingHypothesis}) into Eq,~(\ref{eq-monomers})
and comparing requires us to choose:
$\tau-3 = 2\tau-4$ and
$A = \frac{2-\tau}{1 - \sum_{\w\geq 1} \frac{1}{[\w(\w+1)]^\tau}}$.
This seems to straightforwardly determine the spectral exponent to be $\tau=1$
until we realise that $\sum_{\w\geq 1} \frac{1}{[\w(\w+1)]}=1$ resulting in the 
divergence of the amplitude, $A$, for this choice of  $\tau$. This surprising
result suggests that we consider the more general divergence
\begin{equation}
\label{eq-logarithmicCorrection}
F(x) \sim x^{-1}\, \left[ \ln (1/x)\right]^\rho\hspace{0.25cm}\mbox{as $x\to 0$}.
\end{equation}
where we have introduced a logarithmic spectral exponent, $\rho$, with which
one may hope to cancel the divergence we have just encountered. The tail of 
the wave spectrum then has the form
\begin{equation}
\label{eq-spectrumTail}
N_\w(t) = \frac{A}{s(t)}\,\frac{1}{\w}\,\left[ \ln\left(\frac{s(t)}{\w}\right)\right]^\rho \hspace{0.25cm}\mbox{for $\w \ll s(t)$}.
\end{equation}
Setting $k=1$ in this formula gives the asymptotic form of of $n_1(t)$. 
Substituting these formulae into Eq.~(\ref{eq-monomers}) one finds that 
the leading term on the left hand side is of order 
$s(t)^{-2}\,\ln\left[s(t)\right]^\rho$ and the leading order term on the
right hand side is of order $s(t)^{-2}\,\ln\left[s(t)\right]^{2\rho-1}$ 
({\em not} $s(t)^{-2}\,\ln\left[s(t)\right]^{2\rho}$ as one might
 naively expect owing to the cancellation alluded to above). Thus we should
choose $\rho=1$ for the logarithmic spectral exponent so that the asymptotic
form of the scaling function in the decay case is:
\begin{equation}
\label{eq-decayFAsymptotics}
F(x) \sim x^{-1}\, \ln (1/x)\hspace{0.25cm}\mbox{as $x\to 0$}.
\end{equation}
Fig.~\ref{fig-compensatedScaling} shows the numerically obtained scaling
function rescaled according to this formula. The plateau at small $x$
provides strong numerical support for Eq.~(\ref{eq-decayFAsymptotics}).

\begin{figure}[htbp]
\includegraphics[width=0.4\textwidth]{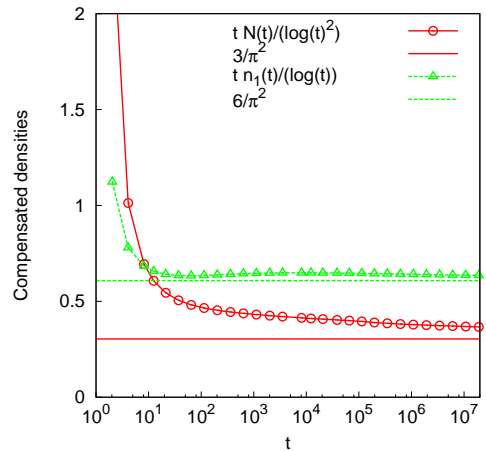}
\caption{Time decay of total wave action, $N(t)$, and primary wave action, 
$\nn_1(t)$, compensated by the theoretically predicted asymptotic decay rates
taken from Eq.~(\ref{eq-densityDecay}) and Eq.(\ref{eq-monomerDecay}) 
respectively. The theoretically expected plateau values are attained to within 
a few percent accuracy.}
\label{fig-densityDecay}
\end{figure}

In principle, one 
should also obtain the amplitude  at this point but this turns out to be
easier using Eq.~(\ref{eq-density}) for $N(t)$ since certain sums which
arise can by computed exactly in that case. The total wave action is
\begin{eqnarray*}
N(t) &=& A s(t)^{-1} \sum_{\w=1}^{s(t)}\w^{-1}\ln\left[\frac{s(t)}{\w}\right]\\
&\approx& A s(t)^{-1} \int_1^{s(t)} \int_{\w=1}^{s(t)}\w^{-1}\ln\left[\frac{s(t)}{\w}\right] \,d\w\\
&=& \frac{A}{2}\frac{\ln\left[s(t)\right]^2}{s(t)}.
\end{eqnarray*}
Substituting this into the left-hand side of Eq.~(\ref{eq-density}), and  
Eq.~(\ref{eq-spectrumTail}) into the right-hand side, and computing the 
leading terms we find the balance
\begin{displaymath}
-\frac{A\,s_0}{2}\,\frac{\ln\left[s(t)\right]^2}{s(t)^2} = 
-\frac{A^2}{2}\, \frac{\ln\left[s(t)\right]^2}{s(t)^2} \, \sum_{\w=1}^\infty \frac{1}{\w^2}.
\end{displaymath}
The sum gives $\pi^2/6$ from which we conclude that 
$\frac{A}{s_0}=\frac{\pi^2}{6}$. Recalling that $s(t)\sim s_0\,t$ this relation
gives us the following nontrivial asymptotic decay laws for the total density
and number of primary waves respectively:
\begin{eqnarray}
\label{eq-densityDecay}
N(t)&\sim& \frac{3}{\pi^2}\frac{(\ln t)^2}{t}\\
\label{eq-monomerDecay}
\nn_1(t)&\sim& \frac{6}{\pi^2}\frac{\ln t}{t}.
\end{eqnarray}
These predictions are validated numerically in Fig.~\ref{fig-densityDecay}.

To conclude, we have used the analogy between three--wave turbulence and 
aggregation--fragmentation processes to study analytically 
the decay kinetics of a simple wave turbulence model. We found that the 
kinetics have non-trivial scaling properties, even in this simple case, which 
differ significantly from the corresponding aggregation process. Our 
results suggest that decaying wave turbulence should be studied in greater
detail than it has been to date.

\section*{Acknowledgements}
We thank the University of Warwick North American Travel Fund for supporting this research.


\end{document}